\documentclass{elsart3}
\usepackage{amssymb} \usepackage{amsmath}
 \usepackage{epsfig} \usepackage{pifont}
\def\albars{\relax\ifmmode{\bar{\alpha}_s}\else{$\,\bar{\alpha}_s${ }}\fi}
\def\as{\relax\ifmmode{\alpha_s}\else{$\,\alpha_s${ }}\fi}
\def\agoth{\relax\ifmmode{\mathfrak A}\else{$\,{\mathfrak A}${ }}\fi}
\def\acal{\relax\ifmmode{\mathcal{A}}\else{${\mathcal{A}}${ }}\fi}
\def\albar{\relax\ifmmode{\bar{\alpha}}\else{$\,\bar{\alpha}${ }}\fi}
\def\eV{{\rm e\kern-0.12em V}}\def\MeV{{\rm M}\eV} \def\GeV{{\rm G}\eV}
\def\MSbar{\relax\ifmmode\overline{\rm MS}\else{$\overline{\rm MS}${ }}\fi}
\begin{document}
\begin{frontmatter}
 \title{Nonpower \ Expansions \ for \ QCD \ Observables \
 at \ Low \ Energies}
 \author{Dmitry Shirkov}
 \ead{shirkovd@thsun1.jinr.ru}
\address{Bogoliubov Lab. of Theor. Physics, JINR,
Dubna, 141980, Russia}

\begin{abstract}
 A comprehensive review is presented of the progress made
 in further developing the ghost-free Analytic Approach to
 low-energy QCD since ``QCD-97" meeting. It is now
 formulated as a logically closed ``Analytic Perturbation
 Theory" algorithm. Its most essential feature is nonpower
 functional expansions for QCD observables.
\end{abstract}

\end{frontmatter}

\section{Intro}\label{s1}
 The claim is that current practice to use expansions for
 QCD observables in powers {\it of the same} effective
 coupling function \albars, i.e., $[\albars(Q^2)]^k\,,
 \,[\albars(s)]^k \,$ or $\,[\albars(r)]^k\,,$ in different
 pictures ($Q^2\,, s\,, r\, ...$) as, e.g., it is implicitly
 recommended by PDG\cite{pdg}, is neither based theoretically,
 nor adequate practically to low-energy QCD. \par
 Instead, one should use {\sf non-power functio\-n\-al
 expansions} with non-power sets of functions
 $\left\{\acal_k(Q^2)\right\}\,, \left\{{\mathfrak A}_k(s)
 \right\}\,$ and $\left\{\aleph_k\right\}\,,$ related by
 some suitable integral transformations.\par

   To comment this, consider a class of homogeneous
 {\large\bf L}inear {\large\bf I}ntegral {\large\bf
 T}ransformations (LITs)
  \begin{equation}\label{lit}
 f_E\to f_i(x_i)= \int K_i\left(\frac{x_i} {Q^2}\right)
 f_E(Q^2)\frac{dQ^2}{Q^2} \,,\end{equation}
 that describe transitions from the Euclidean picture to
 Minkowskian and Distance ones:
 $$ f_E(Q^2)\to f_M(s)\,;\, f_E(Q^2)\to f_D(1/r^2)\,.$$

 Being applied to renormalisation-invariant functions they
 could relate, in particular, effective couplings properly
 defined in these pictures
 \begin{equation}\label{al-trans}
 \alpha_E(Q^2)\to\alpha_M(s)\,; \quad
 \alpha_E(Q^2)\to\alpha_D(r)\,.\end{equation}

   Meanwhile, generally, LITs (\ref{lit}) (and reverse
 transformations) can be used only within a class of functions
 $\,f_i\,$ that are free of nonintegrable singularities.
 For example, a common expression for QCD effective coupling
 \albars with powers of $\,L=\ln (Q^2/\Lambda^2)\,$ in
 denominators (like eq.(9.5) in ref.\cite{pdg})
 is not suitable here.  Due to this, in our further
 discussion we use some adequate models for \albars, free
 of unphysical singularities.\par

  Moreover, LITs change the form of observable expansion;
 being applied to powers of some $\,\albar(Q^2)\,,$ they
 yield expressions that are not powers of the
 $\,\albar(Q^2)\,$ image. They transform expansions in
 coupling powers $\albar^k\,$ into expansions over non-power
 sets $\left\{\acal_k(Q^2)\right\}\,,\left\{{\mathfrak A}_k
 (s) \right\}\,$ and $\left\{\aleph_k(r^{-2}\right\}\,.$

  We give explicit examples of these sets and construct
 non-power expansions for observables that demonstrate
 quicker convergence and reduced renorm-scheme sensitivity.
 \vspace{-3mm}

 \section{Different pictures in HEP}  
 \subsection{Various HEP representations}
 An essential property of any successive theoretical scheme
 for physical phenomena is the possibility to use various
 pictures for their description. In HEP, these are
 space-time, energy--momentum, eikonal and some others.
 Below, we discuss the momentum-transfer (Euclidean part of
 the energy--momentum 4-space defined by
 $Q^2={\bf Q^2}-Q^2_0 \geq 0\,),$ the center-of-mass-energy
 squared (Minkowskian part of the energy--momentum 4-space
 with $\,s= q^2_0-{\bf q^2}\geq 0\,,$) as well as the
 distance (space-like part of space-time
 $\,r^2={\bf x^2}-x^2_0 \geq 0\,$) pictures.\smallskip

 Remind, first, that Euclidean momentum-transfer picture was,
 historically, a natural scene for formulating \cite{q2pict}
 (see, also review Chapter ``Renormalization Group" in
 \cite{book80}) basic renormalisation-group (RG) relations
 and equations. Here, an important feature is that scalar
 propagator amplitudes depending on a real argument
 $\,Q^2\geq 0\,$ are real functions, and their particular
 values can be expressed via real parameters $\,z_i\,$ of
 finite Dyson renormalisation transformations\footnote{
 Analogous possibility is provided by the Euclidean
 space-time picture first used by Dirac \cite{Dirac34} for
 introducing space distribution $\,e(r)\,$ of the electron
 charge.}. Just due to this, we use the {\it Euclidean}
 picture with $\,Q^2 \geq 0\,$ as a launch-platform for
 LITs (1).

 \subsection{Pictures related by integral transformations}
 Transition from an Euclidean function $\,f_E(Q^2)\,$ to the
 Distance one $\,f_D(r^2)\,$ can be performed with an
 appropriate Fourier transformation. For example, in the
 three-dimensional case one has
$$
 f_E(Q^2)\to{\mathbb{F}}[f_E](r^2)=f_D(r^2)\,;$$
 \begin{equation}\label{tranF}
 {\mathbb{F}}[f_E](r^2)\equiv r\int^\infty_0 d Q\,\sin(Qr)\,
 f_E(Q^2)\,.\end{equation}

 At the same time, in current QCD analysis, people use a
 relation between the Euclidean and Minkowskian pictures
 based on definition of the Adler function
 $$
 D(Q^2)=Q^2\,\int^{\infty}_0\frac{d s}{(s+Q^2)^2}\,R(s)\,.$$
  We treat this relation as {\it integral transformation}
\begin{equation}\label{tranD}
 R(s)\to D(Q^2)\equiv{\mathbb{D}}\left[R\right](Q^2)\,,
 \end{equation}
 that transforms a real function $R(s)\,$ of the Minkowskian
 domain (real $s\geq 0\,\,$) to a real function $\,D(Q^2)\,$
 of the Euclidean one (real $\,Q^2\geq 0\,$). The reverse
 transformation $\,\mathbb{R=D}^{-1}\,$ has the form of a
 contour integral.\par

 Both the transformations (\ref{tranF}) and (\ref{tranD})
 are consistent with RG--invariancy and are non--consistent
 with existence of unphysical pole at $\Lambda^2\,.$
 \vspace{-3mm}

\section{Ghost-free APT construction} 
\subsection{Ghost-free QCD coupling in Minkowskian}
 In the early 80s, by summing the $\,\pi^2$--terms arising
 in the course of logs branching
  $$
 \ell\equiv\ln\left(\frac{Q^2}{\Lambda^2}\right) \to
 \ln\left(\frac{s} {\Lambda^2}\right)-i\pi\,\equiv L-i\pi $$
 under the transition $\,Q^2\to -s\,$ Radiushkin, Krasnikov
 and A. Pivovarov  obtained remarkable expressions for
 Minkowskian QCD coupling and its ``effective powers". This
 operation $\,\sum_{\pi^2}\,,$ in the one-loop case for
 $\,\albar_s^{(1)}= 1/\beta_0\ell\,,$ results in the
 explicit form\cite{rad82}:
\begin{equation}\label{agoth1}
 \albars^{(1)}\to\alpha_M^{(1)}(s)= \frac{1}{\beta_0\pi}
 \arccos\frac{L}{\sqrt{L^2+\pi^2}}\,;\end{equation}
  $$
 \left[\alpha_M^{(1)}(s)=\agoth_1^{(1)}(s)\right]_{L>0}=
 \frac{1} {\beta_0\pi}\arctan\frac{\pi}{L}\, $$
 and for square and cube of $\albars^{(1)}\,$ in
 expressions\cite{krapiv82}
\begin{equation}\label{agoth23}
 \frac{1}{\beta_0^2\ell^2}\to \agoth_2^{(1)}(s) =\frac{1}
 {\beta_0^2\left[L^2+\pi^2\right]}\,, \end{equation}
$$
 \left(\albars^{(1)}\right)^3\to \agoth_3^{(1)}(s)=
 \frac{L}{\beta_0^3\left[L^2+ \pi^2\right]^2}\,,$$
 which {\sf are not powers} of $\alpha_M^{(1)}(s)\,.$ As
 it was noticed later\cite{ms97}\footnote{See also
 Ref.\cite{bb-95}.}, they are ghost-free.

\subsection{Ghost-free QCD coupling at Euclidean}  
  In the Euclidean momentum-transfer domain analogous
 ghost-free construction for QCD coupling and observables was
 proposed\cite{prl97} in the mid-90s. This {\it Analytic
 approach} was based on imperative of analyticity in the cut
 $\,Q^2$--plane. Its algorithm, first realized \cite{trio59}
 in QED, combines RG invariance the K\"allen--Lehmann
 spectral representation with--- see, Sections 53 and 54 in
 \cite{book80} and the first of eqs.(\ref{kl}) below.

 In the simplest case of 1-loop QCD coupling, the
 analyticization operation $\mathbb{A}\,$ yields analytic
 Euclidean QCD coupling
 $$
 {\mathbb{A}}\left[\frac{1}{\ell}\,\right]{\Rightarrow}
 \beta_0\alpha^{(1)}_E(Q^2)=\frac{1}{\ell}-\frac{\Lambda^2}
 {Q^2-\Lambda^2}\ ; \ \ell= \ln\frac{Q^2}{\Lambda^2}\,\,.$$

 Analogous expressions with ``subtracted" high-order pole
 singularities are valid for powers of $\albars^{(1)}$;
 e.g., for coupling squared
 $$
 \frac{1}{\ell^2} {\Rightarrow} \frac{1}{\ell^2}+
 \frac{Q^2\Lambda^2} {\left(Q^2-\Lambda^2\right)^2}=
 \beta_0^2\,\acal_2^{(1)}
(Q^2){\neq}\left(\beta_0\alpha_E^{(1)}\right)^2\,. $$

  By construction, new \ analytic \ functions \\
 $\left\{\acal_k\right\}= \acal_1(\equiv\alpha_E)\,,
 \acal_2\,, ... :$ \\
$\bigstar$ are free of unphysical singularities;\\
$\bigstar$ are finite in the IR limit with 
\begin{equation}\label{IRlimit}
 \,\alpha_E(0)=1/\beta_0\,,\quad \acal_{k\geq 2}(0)=0\,;
\end{equation}
$\bigstar$ include {non-perturbative structures};\\
$\bigstar$ in the weak-coupling UV limit $\to$ $\albars^k$
$\sim [\ell]^{-k}\,.$\smallskip

  These properties remain valid for higher-loop cases. In
 particular, IR limiting values (\ref{IRlimit}) provide the
 basis for remarkable stability of the $\acal_k\,$ functions
 behavior in the low-energy domain. On Fig.1, one can see
 that while the 2-loop curve\footnote{For analytic
 expressions and numerical tables of 2-loop functions see
 papers \cite{prl97,2loop}.} for $\,\alpha_E(Q^2)\,$ differs from
 the 1-loop one in the few \GeV \ region by 10-20\%\%, the
 3-loop approximation practically coincides with the 2-loop
 curve, the difference being of the order of 2-3 per cent.
 The same is true for Minkowskian coupling.
 This, in turn, yields low sensitivity with respect to
 change of renorm-scheme (RS).

 \begin{figure}[th] \unitlength=1mm
   \begin{picture}(80,65)                                   %
 \put(-8,0){\epsfig{file=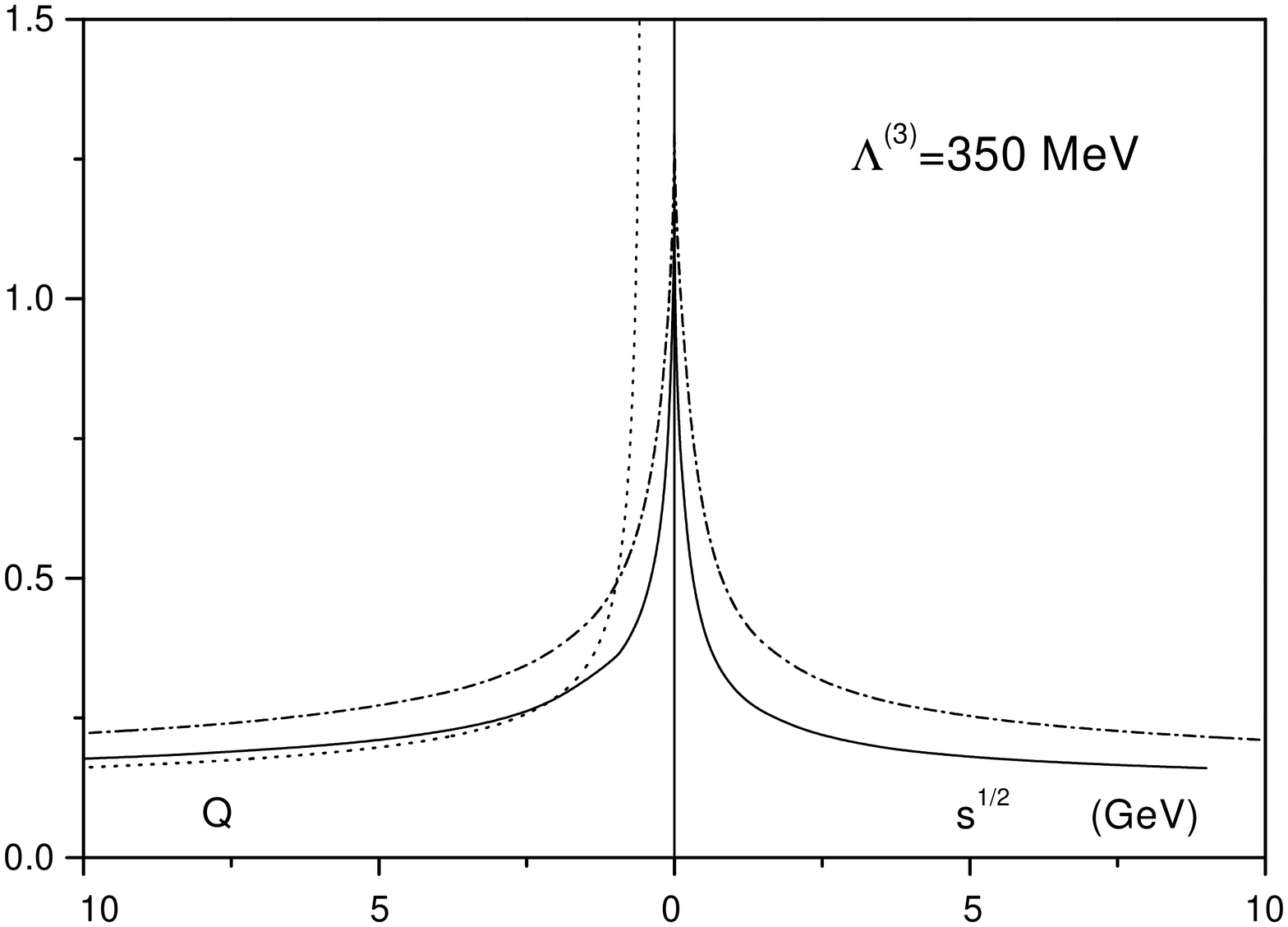,width=8.7cm}}
  \put(37.2,53){$\bullet$}
 \put(25,53){\small $\alpha_E(0)=$}
 \put(41,53){\small $\alpha_M(0)$}
\put(22,43){\small $\bar\alpha_s^{(2)}(Q^2;3)$} %
\put(14,22){\small $\alpha_E^{(1)}(Q^2;3)$}
\put(28,14){\small $\alpha_E^{(2,3)}$}
 \put(44,23){\small $\alpha_M^{(1)}(s;3)$}
 \put(42,13){\small $\alpha_M^{(2,3)}$}
\end{picture}
\centerline{ \parbox{7.1cm}{
 \caption{\sl\footnotesize Space-like and time-like
 global analytic couplings in a few \GeV \ domain
 with $\,\Lambda^{(f=3)}=350 \,\MeV\,.$
 Here, $\,\alpha_{E,M}\,$ in 1-loop case are drown by
 dash-and-dotted lines and for 2-- and 3-- loop cases
  by solid curves. } \label{fig1} }}  \end{figure}

\subsection{Minkowskian vs Euclidean; the APT}
   As it was noticed later \cite{apt01}, both the sets
 of Euclidean and Minkowskian functions are connected by LITs
 $\,\acal_k(Q^2)= \mathbb{D}\left[\agoth_k \right]\,,\,
 \agoth_k(s)= \mathbb{R}\left[\acal_k \right]\,.$ In the
 1-loop case \cite{ms97}
$$ \mathbb{D}\left[\frac{1}{\pi}\arccos\frac{L}
{\sqrt{L^2+\pi^2}}\right]=\frac{1}{\ln(Q^2/\Lambda^2)}\,
-\,\frac{\Lambda^2}{Q^2-\Lambda^2} \,,$$
$$\mathbb{R}\left[\frac{1}{\ln^2(Q^2/\Lambda^2)}
+\frac{Q^2\Lambda^2}{\left(Q^2-\Lambda^2\right)^2}\right]
=\frac{1}{L^2+\pi^2} \,.$$
 Accordingly, for an Euclidean RG-invariant function
 $\,I_E(Q^2)\,$ -- initially expandable in powers of the
 coupling constant \as -- there appears a relation with its
 Minkow\-s\-kian image $\,I_M(s)=\mathbb{R}[I_E]\,$
\begin{equation}\label{7}
\mathbb{D}\left[I_M=\sum_k i_k\,\agoth_k\right] \,=
\,I_E(Q^2)=\sum_k i_k\,\acal_k(Q^2)\,.\end{equation}

  Thus, starting with common RG-improved perturbation
 expansion for an Euclidean function $\,I_{\rm pt}(Q^2,\as)\,
 ,$ after applying operation of analyticization
 $\,\mathbb{A}\,,$ and of $\,\pi^2\,$--summation
 $\,\sum_{\pi^2}\,$ we arrive at non-power functional
 expansions related by (\ref{7}) with numerical
 coefficients $i_k\,.$ The whole construction has been named
 \cite{ms97,apt01,sh1epj} the {\it Analytic Perturbation Theory}
 (APT). It was duly elaborated\cite{apt01} for real QCD
 with various numbers $\,f\,$ of active quarks. The
 logic of its structure is presented in Fig.2.
\begin{figure}[th] \unitlength=1mm
 \begin{picture}(80,60 )(-10,-35)                                   %
\put(-15,-84){\epsfig{file=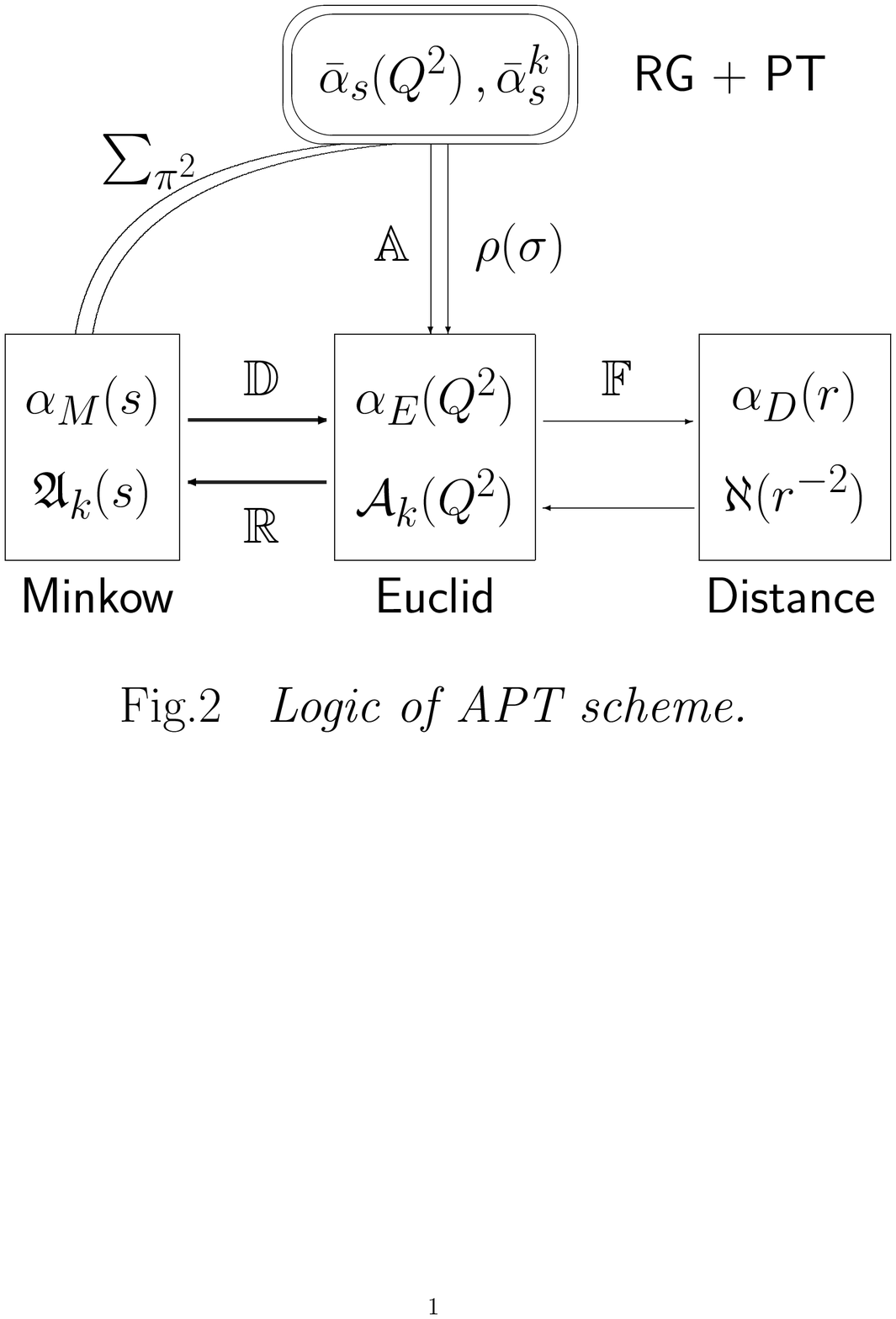,width=8.5cm}}
\end{picture}
 \end{figure} \vspace{-5mm}

\subsection{ Sketch of the global APT algorithm} 
 The most elegant form of the APT formalism uses a
 spectral density $\,\rho(\sigma)\,$ taken from perturbative
 input. Then all involved functions in the above-mentioned
 pictures look like
\begin{equation}\label{kl}
\acal_k=\frac{1}{\pi}\int\limits_{0}^{\infty}
\frac{\rho_k(\sigma)\,d\sigma}{\sigma+Q^2}\,,\ \ 
\mathfrak{A}_k=\frac{1}{\pi}\int\limits^{\infty}_{s}
\frac{d \sigma}{\sigma}\,\rho_k(\sigma)\,,
\end{equation}
 $$
 \aleph_k\left(\frac{1}{r^2}\right)=\int\limits^{\infty}_{0}
 \frac{\rho_k(\sigma)\, d\sigma}{\sigma}\,\left(1-
 e^{-r\sqrt{\sigma}}\right)\,.$$

 In the 1-loop case,
\begin{equation}\label{rhoAPT}
 \rho^{(1)}_1=\frac{1}{\beta_0\left[L_{\sigma}^2+
 \pi^2\right]}\,;\,\quad L_{\sigma}=
 \ln\left(\frac{\sigma}{\Lambda^2}\right)\,\end{equation}
 with higher spectral functions expressed via
  $\,\rho^{(1)}_1\,$ by  simple iterative relation
\begin{equation}\label{iter1}
 k\,\beta_0\,\rho_{k+1}^{(1)}(\sigma)=
 -\frac{d\,\rho_k^{(1)}(\sigma)}{d\,L_{\sigma}}\,,
 \end{equation}
 that implies analogous iterative relations for
 $\,\acal_k\,$ and $\,\agoth_k\,.$
 As it has been noted above, eqs.(\ref{rhoAPT}) and
 (\ref{iter1}) were generalized for a higher-loops case
 with transitions across heavy quark thresholds and were
 successively used \cite{sh1epj} for fitting various data.\par

  Let us emphasize an {\it important ansatz} built-in the
 presented APT construction: {\it spectral functions, like
 (\ref{rhoAPT}), are defined on the basis of RG-improved
 perturbation theory. Due to this, they contain no
 additional parameters and non-perturbative elements.}
 Hence, this construction can be considered as a ``minimal
 APT scheme". \par

 Here, it can be added that some detail of the APT
 construction in the distance picture was recently
 considered in our paper Ref.\cite{dv03}. \vspace{-4mm}

\section{Non-power expansions for observables}
 To summarize APT for observables, we state that instead
 of usual expansions in powers of the same universal QCD
 effective \albars function {\normalsize
 $$
 \left\{\albars^k(Q^2,\as)\right\}  , \
 \left\{\albars^k(s, \as)\right\} , \
 \left\{\albars^k\left(1/r^2,\as \right)\right\}\,$$ 
 one should use  "perturbatively motivated" expansions
 in terms of non-power sets 
$$
 \left\{\acal_k(Q^2,\as)\right\}\,,\ \left\{
 {\mathfrak A}_k(s,\as)=\mathbb{R}[\acal_k](s)\,\right\}\,,$$
$$ \left\{\aleph_k(1/r^2,\as) =\mathbb{F}[\acal_k](1/r^2)\,
\right\}\, $$ 
 with some of them being non-analytic at $\as=0\,.$
\vspace{-1mm}

\subsection{Properties of non-power functions}
  Review now the main properties of nonpower sets
 $\left\{\acal_k\right\},\left\{\agoth_k \right\}\,,
 \left\{\aleph_k \right\}\,.$ Qualitatively, all three are
 similar: \par

 I.\ They consist of ghost-free  functions \par
 II. \ First functions, new couplings,
{\bf $\alpha_E,\,\alpha_M,\,\alpha_D$ } \\
{$\diamondsuit$} are  monotonic; \\
{$\diamondsuit$} In the IR limit, they are finite
 {\normalsize $=1/\beta_0\simeq 1.4\,$};\\
{$\diamondsuit$} In the UV limit \ \
$\,\sim 1/\ln x\/\sim \albars(x)\,.$ \smallskip

 III.  All other functions \ $(k\geq 2)\,:$ \\
{$\heartsuit$} start from zero {\normalsize
 $\acal_k(0),{\mathfrak A}_k(0),\aleph_k\,(0)=0;$ }\\
 $\heartsuit$ in the UV limit $\sim 1/(\ln x)^k\,\sim
 \albars^k(x)\,.$\\
 {$\heartsuit$} 2nd ones, $\acal_2\,,\,{\mathfrak A}_2\,,
 \aleph_2\,$ \ obey max at $ \sim \Lambda^2\,.$\\
{$\heartsuit$} Higher ones $\,k\geq 3\,$ oscillate at
 $\sim\Lambda^2$ and obey $\,k-1\,$
 zeroes\footnote{On mathematical nature of these sets
 see papers \cite{lmp99,tmp99}.}.\par
 Just two last properties play an essential role is the
 RS-insensitivity and in better convergence of expansions
 in the low-energy region.
\addtocounter{figure}{1}
 \begin{figure}[th] \unitlength=1mm
   \begin{picture}(70,65)
 \put(-8,1){\epsfig{file=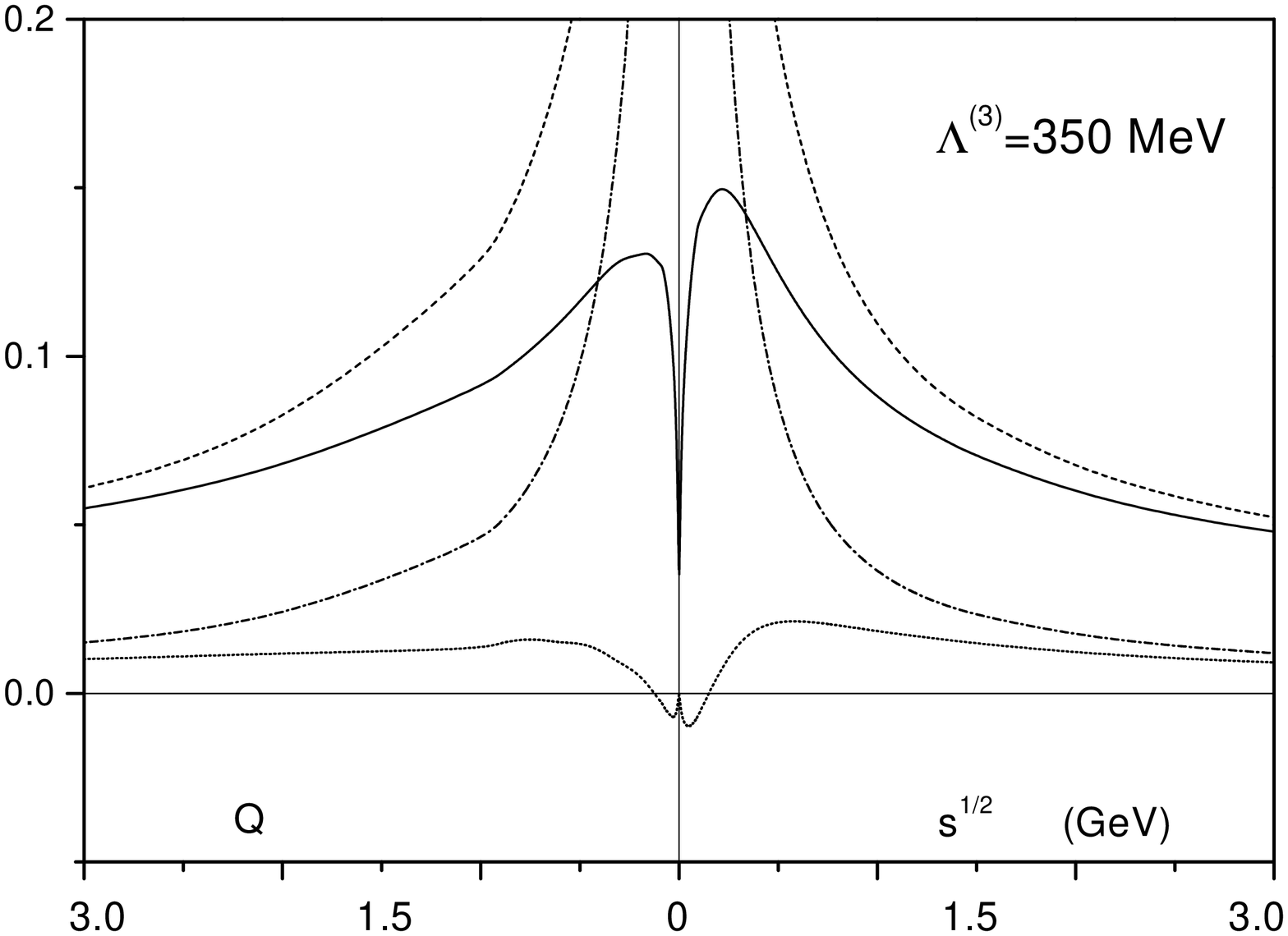,width=8.7cm}} %
 \put(23,52){\small $\alpha_E^2$}%
  \put(18,32){\small\bf $\acal_2$}%
  \put(29,30){\small $\alpha_E^3$}%
 \put(30,25){\small\bf $\acal_3$}
 \put(52,42){\small $\alpha_M^2$}%
 \put(55,30){\small ${\mathfrak A}_2$}%
 \put(46.5,32){\small  $\alpha_M^3$}
    \put(41,25){\small ${\mathfrak A}_3$}%
  \end{picture}
 \centerline{ \parbox{7.3cm}{
 \caption{\sl\footnotesize  ``Distorted mirror symmetry"
 for global expansion functions $\{\acal_k(Q^2)\}\,$ and
 $\left\{{\mathfrak A}_k(s)\right\}.$ All the curves here
 correspond to exact two--loop solutions expressed in terms
 of Lambert function. One can see that 3rd functions are
 oscillating in the low-energy region. There, they are
 much less than third powers (dot-and-dash lines) of
 related coupling functions.} \label{fig3}}}
 \end{figure}

   It is curious that the mechanism of liberation of
 singularities is quite different: for $\acal_k\,,$ in the
 space-like domain, it involves subtraction of
 nonperturbative, power in $Q^2\,,$ structures. Meanwhile,
 in the time-like region, it is based only on resummation of
 ``$\pi^2\,$-- terms". Graphically, {\it nonperturbative
 analyticization\/} in the $Q^2$--channel looks like a
 slightly distorted reflection (as $Q^2\to s=-Q^2$) of the
 perfectly perturbative {``$\pi^2\,$--summation"} result in
 the $\,s$--channel. The effect of ``distorting mirror"
 (first discussed by Milton-Solovtsov \cite{ms97}) is
 illustrated in Figs. 1 and 3.\vspace{-2mm}

\subsection{Better convergence and RS insensitivity}
 Due to a quite different from the usual oscillation behavior
 of higher (with $\,k\geq 3\,$) functions, the APT nonpower
 expansions 
\begin{gather*}
d_{\rm an}(Q^2)=d_1\,\alpha_E(Q^2)
 +d_2\,\acal_2(Q^2)+ d_3\,\acal_3+... \\
 r_{\pi}(s)= d_1\alpha_M(s)+d_2\,{\mathfrak A}_2
(s)+d_3\:{\mathfrak A}_3(s)+...
\end{gather*} 
 obey better convergence in the hadronization region. We
 demonstrate this by a Table from paper \cite{sh1epj}.
 All results are given in the \MSbar scheme.\par

 As it follows from the comparison between the 3rd columns
 for usual (PT) case and for the APT one, the relative
 contribution of the 3rd ($\,\sim \albars^3\,$) term
 drastically diminishes; it is not any more of the same
 order of magnitude as the 2nd term. \par

\begin{center}
\begin{tabular}{|l||c|c|c||r|r|r|}
\multicolumn{7}{c}{\sf Table: Contributions in \%\% of 1-- ,
 2-- }\\
 \multicolumn{7}{c}{\sf and 3--loop terms into observables}\\
 \hline 
\multicolumn{1}{|l||}{\slshape \phantom{aa}
{\huge $\phantom{.}_{\rm Process}$ \ \ }} &
\multicolumn{3}{c||}{\slshape {\small PT}}
 & \multicolumn{3}{r|}{\slshape{\small APT}
 \phantom{aaaa}} \\ \cline{2-7}%
 & 1st & 2nd & 3rd &  1st & 2nd & 3rd \\ \hline\hline
{GLS \ Sum \ Rules} & 65 & 24 &\large\bf{11} & 75 & 21 & \large\bf{4} \\ \hline
{Bjorken Sum Rules} & 55 & 26 &\large\bf{19} & 80 & 19 & \large\bf{1} \\ \hline\hline
Incl. V.\, $\tau$-decay & 55 & 29 &\large\bf{16} & 88 & 11 & \large\bf{1}\\ \hline
{\normalsize 10\GeV\, $e^+e^- \to $ hadr} & 96 & 8 & -4 & 92 & 7 & \bf{.5}\\ \hline
$Z_o \to$ hadrons & 98.6 & 3.7 & -2.3 & 96.9 & 3.5 & -\bf{.4} \\ \hline
\end{tabular} \end{center}
\medskip

  In turn, this leads to essential reducing of dependence of
 the results on change of the renormalisation-scheme
 prescription. For further numerical examples of reduced RS
 sensitivity see papers \cite{pl98,mss98,mss99}. One more fresh
 illustration is provided by recent analysis of the pion form
 factor \cite{bpss04} presented by N.Stefanis at this meeting.

\subsection{Issue of the APT non-uniqueness}\vspace{-2mm}

 The described minimal APT scheme obeys one specific feature:
 it does not contain any new parameters. Like in usual QCD
 case, there is only one adjustable parameter; it can
 be chosen as $\,\Lambda_{(f=5)}\,$ or $\,\albars(m_{Z_0})\,.$
 Aesthetically, this feature looks very nice. In particular,
 as a thorough numerical analysis reveals, within the APT
 framework with its two different coupling functions
 $\,\alpha_M\,$ and $\,\alpha_E\,,$ it turns out to be
 possible to correlate a great bulk of QCD data more
 successively\cite{sh1epj} (with smaller $\,\chi^2\,$ value)
 than by a standard algorithm with the only \albars -- see
  the paper \cite{ziggi4}. \par

  Meanwhile, there exist nonperturbative evidence that in
 the 1 \GeV \ region QCD coupling could reach values close to
 \ 0.6-0.9,\cite{al-1gev} (see also our review \cite{tmp02})
  which is impossible for the minimal APT effective couplings
 at realistic $\,\Lambda\,$ values. \par
  In APT, there are several possibilities for inserting
 adjustable ``low-energy parameters". The simplest -- from
 physical point of view -- one, is to attribute some effective
 low-energy mass to partons. This idea is not a new one ---
 see, e.g., refs.\cite{part-mass}.
 In the APT context this trick was successively used by
 Baldicchi and Prosperi \cite{bp02} in calculating mass
 spectrum of light mesons. \par

   Another way is to insert some non-perturbative structure
 into spectral density $\rho(\sigma)\,$ in eqs.(\ref{kl}).
 An example of this kind can be found in Refs.\cite{nester}.
 \medskip

 \centerline{\bf Acknowledgements}\smallskip

  It is a pleasure to express my gratitude to Drs. A.Bakulev,
 S.Mikhailov, X.Papavassiliu, I.Solovtsov and N. Stefanis for
 stimulating discussions.
  This research was supported by RFBR grant No. 02-01-00601
  and by Russian Scientific-School Grant No. 2339.2002.2.

\end{document}